\documentstyle[12pt,twoside,cmp209,epsfig]{article}
\newcommand\bsigma{{\bf S}}
\newcommand\bnabla{\mbox{\boldmath $\nabla$}}

\newcommand{\be}{\begin{equation}}
\newcommand{\ee}{\end{equation}}
\newcommand{\<}{\langle}
\renewcommand{\>}{\rangle}
\renewcommand{\r}{{\bf r}}
\newcommand{\q}{{\bf q}}

\newcommand\x{{\bf x}}
\newcommand\k{{\bf k}}

\title[Nematic phase transitions in two-dimensional systems]
	{Nematic phase transitions in two-dimensional systems}
\author[Berche and Paredes]
	{B. Berche\refaddr{*,**} and R. Paredes\refaddr{**}
        }
\addresses{
        \addr{*} {Groupe M, Laboratoire de Physique des Mat\'eriaux, }
        {UMR CNRS 7556,}\\
        {Universit\'e Henri Poincar\'e,  Nancy 1,}\\
        {F-54506 Vand\oe uvre les Nancy Cedex, France}\\
        \addr{**} {Centro de F\'\i sica,}\\
	{Instituto Venezolano de Investigaciones Cient\'\i ficas,}\\
	{Apartado 21827, Caracas 1020A, Venezuela}\\[2mm]
        {\footnotesize\tt berche@lpm.u-nancy.fr}\\
        {\footnotesize\tt rparedes@ivic.ve} \\[-3mm]
}
\begin{document}

\maketitle

\begin{abstract}
Simulations of nematic-isotropic transition of liquid crystals
in two dimensions are  performed using an $O(2)$ vector model
characterised by non linear nearest neighbour spin 
interaction governed by the fourth Legendre polynomial $P_4$.
The system is studied through standard Finite-Size Scaling
and conformal rescaling of density profiles or correlation functions. 
The low temperature limit is discussed in the spin wave 
approximation and confirms
the numerical results, while the value of the correlation function 
exponent at the deconfining transition seems controversial.
\ {Fluctuation phenomena, random processes, and
Brownian motion}, 
      {64.60.Fr}\ {Equilibrium properties near critical points,
critical exponents},
      {75.10.Hk}\ {Classical spin models}
\end{abstract}

\section{Ordering in two dimensions\label{Sec:Intro}}
\begin{figure}[h]
	\vskip-20mm
  \begin{minipage}{\textwidth}
  \epsfig{file=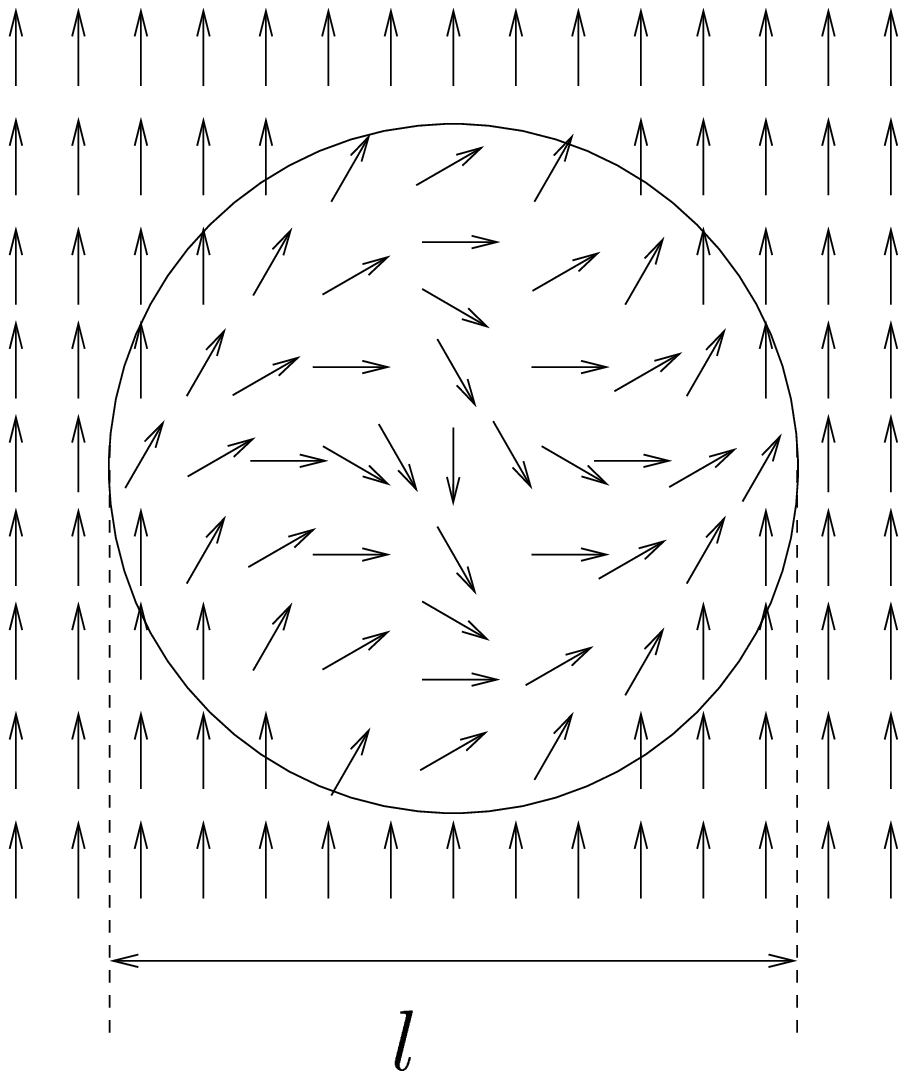,height=0.90\textwidth}\hfill\qquad
  \end{minipage}\vskip-130mm
  \begin{minipage}{\textwidth}
  \ \hfill\epsfig{file=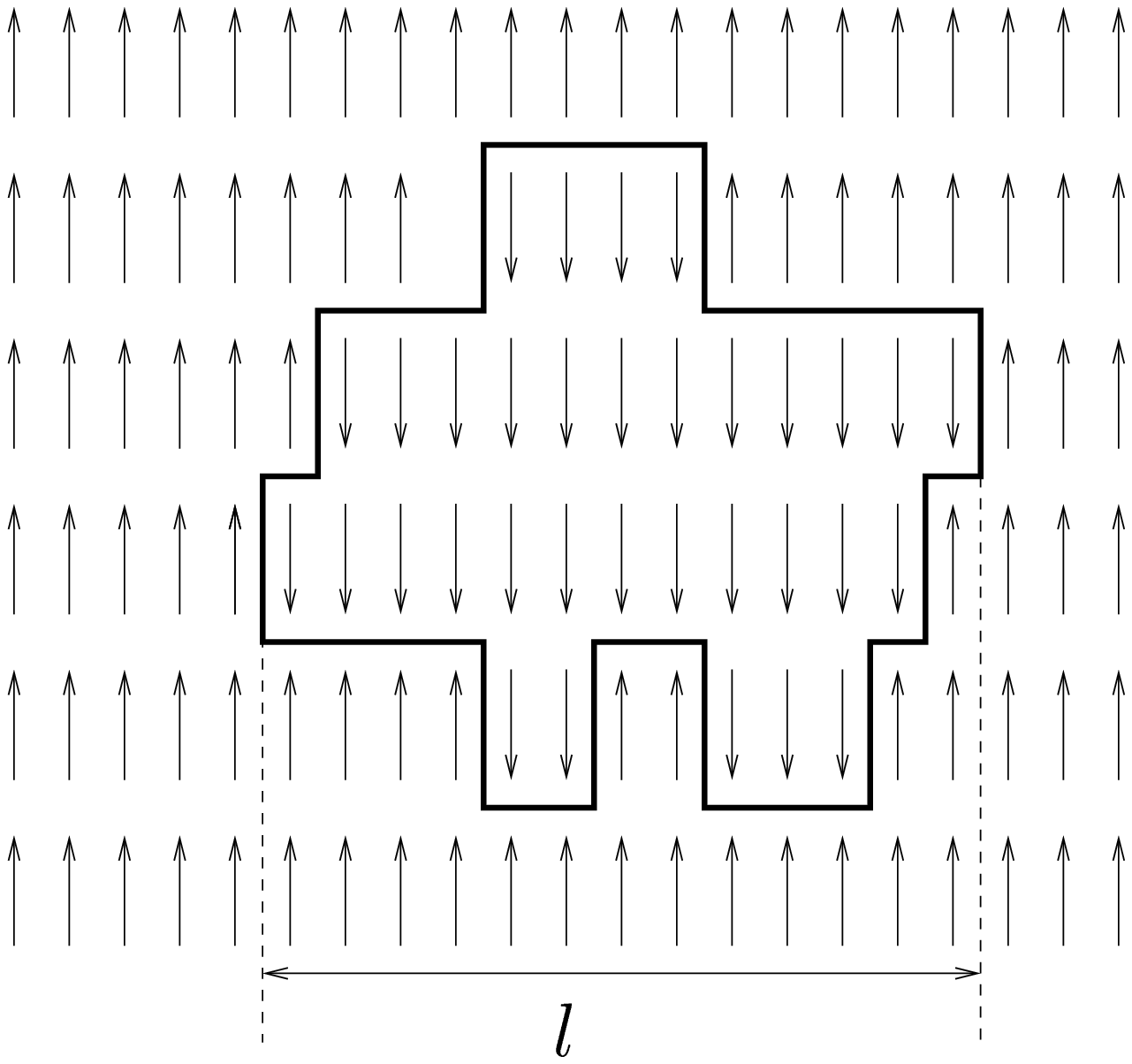,height=0.75\textwidth}\qquad
  \end{minipage}\vskip-45mm
  \caption{Evaluation of the free energy of a 
	disordering droplet in two dimensions in the case
	of continuous symmetry (left) and discrete symmetry (right).}
	\label{fig1}
\end{figure}

In the context of phase transitions, two-dimensional models 
exhibit a very rich 
variety of typical behaviours, ranging from conventional temperature-driven
second order phase transitions (e.g. Ising model) to first-order ones 
(e.g. $q>4-$state
Potts model), with also specific
properties of models with continuous global symmetry which may present
defect-mediated topological phase transitions (e.g. $XY$ model) 
or even no transition at all (e.g. Heisenberg model).
Models of nematic-isotropic 
orientational phase transitions belong to this latter
category of systems displaying a continuous  symmetry.
Ordering in low dimensional systems is likely to be frustrated by
the strentgh of fluctuations. On qualitative 
grounds, consider e.g. how fluctuations develop within the framework
of Landau theory when the temperature 
decreases from the high temperature paramagnetic phase toward the
transition temperature. The response to a localized magnetic field
applied at the origin, $h\delta(\x)$, follows from
Ginzburg-Landau functional minimization of the
free energy $F[m(\x)]=\int d\x (\frac 12am^2(\x)+\frac 14bm^4(\x)
+K|\nabla m(\x)|^2-h\delta(\x) m(\x))$. 
After linearization and Fourier transform, it yields 
\be \xi^{-2}\tilde m(\k) 
+|\k|^2\tilde m(\k)=\frac{h}{2K}
,\ee
where $\xi^{-2}=a/2K$ is the correlation length.
The response $\tilde m(\k)$ is here proportional to the correlation function
Fourier transform $\tilde G(\k)$ and the fluctuations are measured through
\be k_BT\chi=V^{-1}(\<M^2\>-\<M\>^2)=\sum_\k\tilde G(\k).\ee
The latter sum is converted to an integral from $0$ to some cutoff $\Lambda$.
It is diverging with $\xi$ in $1d$ ($\sum_\k\tilde G(\k)
\sim\xi {\rm Arctan}\ \!(\Lambda\xi)$) and in $2d$
($\sum_\k\tilde G(\k)
\sim\ln(\Lambda\xi)$) 
while it is bounded in $3d$ ($\sum_\k\tilde  G(\k)
\sim{\rm const.}$). The divergence in $1d$  prevents any 
long range ordering, while stable ordered state is not forbidden in 
three-dimensional systems. In the intermediate case, due to the logarithmic
diverging behaviour of the intergral it is less obvious to conclude and
more refined analysis is required.
In his famous book on phase transitions, Cardy uses a simplified version
of Peierls argument on the existence of a  
phase transition at finite temperature in $2d$ in the
case of discrete symmetry and extends the argument to continuous symmetry, 
showing that ordered ground state is unstable with respect to thermal 
fluctuations in this latter situation.  Consider a spin system with
only nearest-neighbour interactions $-J\bsigma_i\bsigma_j$ 
and assume that the spins
are represented by classical $n-$component vectors.
An ordered ground state may be
stabilized e.g. by symmetry breaking fields at some boundaries of the system.
The variation of internal energy
when a droplet of typical size $l$ with spins progressively tilted in such a 
way that at the center of the droplet the spins are pointing opposite to the
direction of the field is of order of $O(l^2)\times J|\bsigma|^2(\pi/l)^2$ 
where 
$\pi/l$ is the nearest neighbour spin disorientation. This result
follows from  integration over
the droplet volume $O(l^2)$. 
Considering that the entropy is measured by the number
of possible closed loops of size $O(l)$ in $2d$, the 
entropy of the droplet is estimated as
$k_B\ln\mu^l$ where $\mu< z-1$ ($z$ is the coordination of the lattice),
so that eventually the free energy variation is
$\Delta F(l) =
\pi^2 J|\bsigma|^2-k_BTl\ln\mu$. 
At any non zero temperature, 
increasing the size $l$ of the droplet stabilizes the system and a 
spontaneously 
ordered ground state is thus impossible.
In the case of discrete symmetry, e.g. Ising model,  
the energy balance would be associated
to the interface only and we would get
$\Delta F(l)=
2J|\bsigma|^2l-k_B T l\ln\mu$, 
showing that a transition to a stable ordered ground state is expected at 
a temperature around $k_BT_c=O(2J|\bsigma|^2)$.
The two examples are illustrated in Fig.~\ref{fig1}.
In the latter situation, a conventional phase transition toward an
ordered phase is expected at finite temperature, while such an ordered
phase may only be encountered at zero temperature in the first 
example~\cite{MerminWagner66,Hohenberg67}.
On the other hand, an unconventional phase transition toward a
quasi-long-range ordered state may take place at finite temperature,
as we discuss in the next section.

\section{Two-dimensional electrodynamics and the $XY$ model}
The celebrated $XY$ model, 
\be H=-J\sum_{(\r,\r')}\cos(\theta_\r-\theta_{\r'})\simeq
{\rm const}+ \frac12 J\sum_{(\r,\r')}(\theta_\r-\theta_{\r'})^2,
\label{eq-XY}
\ee
admits a phenomenological description in terms of 
Coulomb gas. This is a well-known description, first given by Berezinski\u\i\  
and
Kosterlitz and Thouless~\cite{Berezinskii71,KosterlitzThouless73,Kosterlitz74}
and it is worth reminding its essential steps here.
Before considering this model, 
imagine a point-charge $q$ located at site $\r_0$ in a two-dimensional
space, $\rho(\r)=
q\delta(\r-\r_0)$. The corresponding Coulomb potential, solution of
Poisson equation $\bnabla^2\phi(\r)=-\varepsilon_0^{-1}\rho(\r)$ 
might be written 
\be\phi(\r-\r_0)=-\frac q{2\pi\varepsilon_0}\ln \frac{|\r-\r_0|}{a}\ee
where $a$ is chosen such that $\phi(a)=0$.
Consider now a spin system living in two-dimensional space and
let call ${\bf u}=\bnabla\theta$ the distorsion field, where
$\theta(\r)$ is the phase field defined by
$\bsigma(\r)=(\cos\theta(\r),\sin\theta(\r))$. Due to the periodicity
of the phase field, the distorsion field should obey the following
relation
$\oint_{{\cal C}(\r_0)}{\bf u}\ \!d{\bf l}=2\pi\times{\tt integer}$ 
where the contour integral
is taken along a counterclockwise closed path around  
the point $\r_0$. Using Stokes theorem,
we may also write
\be
\oint_{{\cal C}(\r_0)}{\bf u}\ \! d{\bf l}=
\int d^2r \ \! \hat{\bf z}(\bnabla\times{\bf u})=
2\pi\times{\tt integer}\ee
which implies
\be 
\bnabla\times{\bf u}=
2\pi\times {\tt integer}\times\delta(\r-\r_0)\hat{\bf z}.
\ee
The general solution of this equation consists of two terms,
${\bf u}=\bnabla\psi-\bnabla\times(\hat{\bf z}\phi)$. The curl of the
first term being identically zero, 
we are led to a Poisson equation for the singular term
$\bnabla^2\phi=
2\pi\times {\tt integer}\times\delta(\r-\r_0)$ 
where the ${\tt integer}\equiv n$ plays the role of the total (topological)
charge enclosed by the
contour ${\cal C}$.
From an energetic point of view, we may consider the kinetic energy
\be
\beta H=\frac 12 K\int  d^2r\ \! |{\bf u}|^2\label{defHKinetic}
\ee
which, up to an inessential constant, 
is the continuous approximation of equation~(\ref{eq-XY}) with 
$K=\beta J$.
After decomposition in $\psi$ and $\phi$ parts, the cross-term vanishes and
we get two independent contributions,
\be
\begin{tabular}{llclc}
$\displaystyle \beta H\!\!\!\!$&$\displaystyle =\!\!\!\!$&
$\displaystyle \frac 12 K\int  d^2r\ \! |\bnabla\psi|^2$
&$\!\!\!\!\displaystyle +\!\!\!\!\!\!$&
$\displaystyle \frac 12 K\int  d^2r\ \! |\bnabla\phi|^2$.\\
&&{\tt spin\ waves}&&{\tt defects}
\end{tabular}
\ee
This is a fundamental relation  which is the basis of the factorization
of the partition function in a spin wave contribution and a defect (vortex)
contribution, 
$Z={\rm Tr} \exp(-\beta H)=Z_{{\rm SW}}Z_{{\rm V}}$. 
In the lattice version, the spin wave contribution to the Hamiltonian
$\beta H_{{\rm SW}}=\frac 12K\sum_{(\r,\r')}(\psi_\r-\psi_{\r'})^2$
is quadratic in Fourier space,
\be
\beta H_{{\rm SW}}=\frac 12K\sum_{\q}|\q|^2\mu^2|\psi_\q|^2
\ee
($\mu$ is the lattice spacing  and $\psi_{\bf q}$ the
Fourier component of $\psi_\r$).
This expression leads to the Gaussian
model which implies~\cite{Rice65} that 
\be\<\bsigma_{\r_1}\cdot\bsigma_{\r_2}\> 
={\rm e}^{
 -{\textstyle \frac 12}\langle(\psi_{\r_1}-\psi_{\r_2})^2\rangle} 
\simeq
|\r_1-\r_2|^{-1/2\pi K}\ee
where a temperature-dependent spin-spin critical exponent
is found, $\eta_{{\rm SW}}=(2\pi K)^{-1}$. 
The low temperature (LT) phase of the $XY$ model is a 
quasi-long-range ordered phase (QLRO), 
or a critical phase. When the temperature increases, vortices 
bounded in pairs produce further
disordering of the system and a faster than linear increase of the exponent
$\eta$ with temperature up to the temperature where the transition takes place.
The role of vortices is understood perturbatively through the calculation
of the effective screened interaction energy between the topological
charges.
We consider a neutral Coulomb gas. Omitting a diverging 
contribution $O(\ln R/a)\times\sum_i n_i$
which would occur otherwise (i.e. for a non-neutral Coulomb gas), 
the vortices contribution to the Hamiltonian reads as
\begin{eqnarray}
\beta H_{{\rm V}}&=&-\frac 12K\int d^2r\ \! \phi\bnabla^2\phi\nonumber\\
&=&-2\pi K\sum_{ij\atop i\not= j} n_in_j\ln\frac{|\r_i-\r_j|}{a}=
\sum_{ij\atop i\not= j}\beta V(\r_i-\r_j)\label{eqHV}
\end{eqnarray}
where $V(\r_i-\r_j)$ is the Coulomb interaction between the charges.
In the perturbative approach, we consider the effective interaction between
charges $n_i$ and $n_j=-n_i$ in the presence of another screening
dipole. The following result comes out~\cite{Nelson02}
\be {\rm e}^{-\beta V_{\rm eff}(\r_i-\r_j)}=
{\rm e}^{-\beta V(\r_i-\r_j)}\Bigl(1+{\rm const}\times y_0^2\times \int
	d r\ \! r^{3}{\rm e}^{-\beta V(r)}+O(y_0^4)\Bigr),\label{eqPerturbation}\ee
where $y_0$ is the fugacity of a charge. The correction term 
$y_0^2\int dr\ \! r^{3-2\pi K}$ diverges at the deconfining transition
$2\pi K_c=4$ where the pairs of charges break. The presence of vortices
increases the disordering of the system which,
below $T_c$,  flows under renormalization to the zero-fugacity limit 
where the spin-wave limit is 
recovered  with a renormalized temperature~\cite{Kosterlitz74}. 
At the transition, the correlation
function exponent takes a universal value
\be \eta_c=\frac{1}{2\pi K_c}=\frac 14.\ee

The Heisenberg model ($O(3)$) on the other hand
has no transition at any finite temperature (asymptotic freedom). 
An intuitive argument called 
{\em escape to the third dimension} is often reported. 
This is the observation than vortices cannot be stable for $n\ge 3$, since
an infinitesimal amount of energy is able to produce a spin-wave-type
excitation which eliminates the localized defect. 
Also a renormalization scheme was proposed by Polyakov to treat the
non linear $\sigma-$model close to the lower critical dimension
$d=2$~\cite{Polyakov75,BrezinZinnJustin76,Amit84,Izyumov88}. The flow of the coupling
constant $g$ (the temperature) under a change of scale $s$ is
governed by the $\beta-$function,
$\beta(g,d)=\left. s\frac {\partial g}{\partial s}\right|_0
	=(d-2)g-(n-2)g^2+O(g^3)$ 
which shows that the model is disordered at any temperature when $n>2$ in
two dimensions, since the coupling always decreases and flows to zero under 
renormalization.

\section{Nematics}
\subsection{Definition of the model and of the observables}
We now come back to liquid crystals,
the molecules of which may be idealized as long neutral 
rigid rods. They are likely to interact through electrostatic
interactions, therefore Legendre polynomials appear
for the description of the orientational transition between a disordered
isotropic high temperature phase and an ordered nematic phase. 
Lattice models of nematic-isotropic transitions capture the essentials
of this extremely simplified description. 
The molecules are represented by $n-$component unit
vectors $\bsigma_\r$ (also called ``spins''), located here on the sites $\r$ 
of a square lattice.
The interaction between molecules is restricted to the nearest neighbour
pairs
$(\r,\r')$,  the radial dependence being kept constant and the angular 
dependence entering through a $k-$th order Legendre polynomial\footnote{Even
order Legendre polynomials guarantee the local $Z_2$ symmetry 
$\bsigma_\r\to -\bsigma_\r$.}, 
$P_k(\bsigma_\r\cdot\bsigma_{\r'})$,
in terms of the scalar product between
$\bsigma_\r$ and $\bsigma_{\r'}$,  $\bsigma_\r\cdot\bsigma_{\r'}$. 
A coupling parameter
$J$ measures  the interaction intensity. One obtains the following
Hamiltonian of a lattice liquid crystal,
\be
	H=-J\sum_{(\r,\r')}P_k(\bsigma_\r\cdot\bsigma_{\r'}).
\label{eq:defH}
\ee
When the value of $k$ in Eq.~(\ref{eq:defH}) is varied,  new features
may be expected, as in the case of  
symmetry-breaking magnetic fields $h_k\cos k\theta$ added to the $XY$ model 
which change the phase diagram as investigated by Jos\'e et 
al~\cite{JoseEtAl77,Nelson02}. 
When the polynomial order $k$ increases, 
one may  expect a qualitative change 
in the nature of 
the transition, like in the case of discrete spin symmetries 
(Potts model)~\cite{DomanySchickSwendsen84,JonssonMinnhagenNylen93}. 
The value $k=2$ was intensively studied. It still corresponds to the $XY$ model
for $O(2)$ spin symmetry, while it leads to the $RP^2$ or 
Lebwohl-Lasher model~\cite{LebwohlLasher73} for $3-$component spin vectors. 
The nature of the transition in this
latter case is still under discussion: an early study of 
Kunz and Zumbach~\cite{KunzZumbach92} reported numerical evidences in favour 
of a topological transition, but
more recently, several authors
argued in favour of the absence of any finite-temperature phase 
transition~\cite{BloteGuoHilhorst02,CaraccioloPelissetto02}, 
like in the Heisenberg case. 
Our own previous
contributions~\cite{FarinasParedesBerche03,FarinasParedesBerche04} support
the first scenario with QLRO at low temperature like in the $XY$ model, 
but one cannot exclude a finite - but extremely large -
correlation length which exceeds the maximum size available in numerical
simulations. 
A recent study reported  extremely convincing new evidences in favour 
of a topological transition~\cite{DuttaRoy04}, the transition being driven
by topologically stable point defects known as $\frac 12$ disclination points.
Eventually, in the large$-n$ limit, there is a proof of asymptotic 
freedom for values
of $k$ (in the interaction term $(1+\cos\theta)^k$) 
which do not exceed a critical 
$k_c\simeq 4.537...$~\cite{CaraccioloPelissetto02}.
Above this value the transition becomes of first order, 
a result which does not violate 
Mermin-Wagner-Hohenberg theorem, 
since the correlation length is finite
at the transition.
For finite value of $n$, the question of the nature of the transition
at high $k$ is still a challenging problem, although there is a rigourous 
proof that the transition becomes of first-order for large enough values of $k$
for arbitrary $n\ge 2$~\cite{vanEnterShlosman02,vanEnterShlosman05}. 
This observation suggests to inspect the effect of a higher value of $k$ also
for the $O(2)$ model.
In the context of orientational
transitions in liquid crystals, Legendre polynomials rather than $\cos^k\theta$
interactions are introduced, and we are led to the Hamiltonian of 
Eq.~(\ref{eq:defH}).

In this report,  
we consider the behaviour of an Abelian
spin model ($O(2)$ rotation group) with
$P_4-$like spin interactions. We will refer to it as the $P_4\ O(2)$ model
for simplicity. 
For $2-$component vectors in a disordered phase, $\<\cos^2\theta\>=\frac 12$
and $\<\cos^4\theta\>=\frac 38$. In order to keep the same symmetry in
the interaction than in the $P_4\ O(3)$ model already considered in the 
literature~\cite{MukhopadhyayPalRoy99,PalRoy03}, but to 
normalise it between 0 and 1 in the limits of 
completely disordered and completely ordered phases respectively, we 
modify slightly the Hamiltonian, considering pair interactions of the form
$Q_4(x)\equiv AP_4(x)+{\rm const}=\frac8{55}(35x^4-30x^2+\frac{15}8)$. The
corresponding Hamiltonian is thus defined by
\be
	H_{P_4\ O(2)}=
	-J\sum_{(\r,\r')}Q_4(\bsigma_\r\cdot\bsigma_{\r'}),
	\label{eq:defHO2}
\ee
with 
$\bsigma_\r=(S_\r^x,S_\r^y)$, $|\bsigma_\r|=1$.
A qualitative description of the transition  is
provided by the temperature behaviour of the energy density, the specific
heat, the order parameter  and the susceptibility.
The internal energy is defined from the thermal average of the Hamiltonian 
density, 
$
	u(T)=(dL^d)^{-1}\<H\>
$
and the specific
heat follows from fluctuation dissipation theorem, 
$
	C_v(T)=(L^dT^2)^{-1}(\<H^2\>-\<H\>^2).
$
Brackets denote the thermal average.
The definition of the scalar order parameter (sometimes called nematisation) 
is deduced from the local
second-rank order parameter tensor, 
$ M^{\alpha\beta}(\r)=S_\r^\alpha
S_\r^\beta-\frac 12\delta^{\alpha\beta}$.
After space averaging,
the traceless tensor  $L^{-d}\sum_\r M^{\alpha\beta}(\r)$ 
admits two opposite eigenvalues $\pm\frac 12\eta$ corresponding to eigenvectors
${\bf n }_+$ and ${\bf n}_-$.
The order parameter density is defined after thermal averaging by
$M_2(T)=\<\eta\>$. 
The associated susceptibility is defined by the fluctuations of the order
parameter density, 
$\chi_{M_2}(T)=\frac{4L^d}{k_BT}(\<\eta^2\>-\<\eta\>^2)$.

\begin{figure}[h]
  \centering
  \begin{minipage}{\textwidth}
  \epsfig{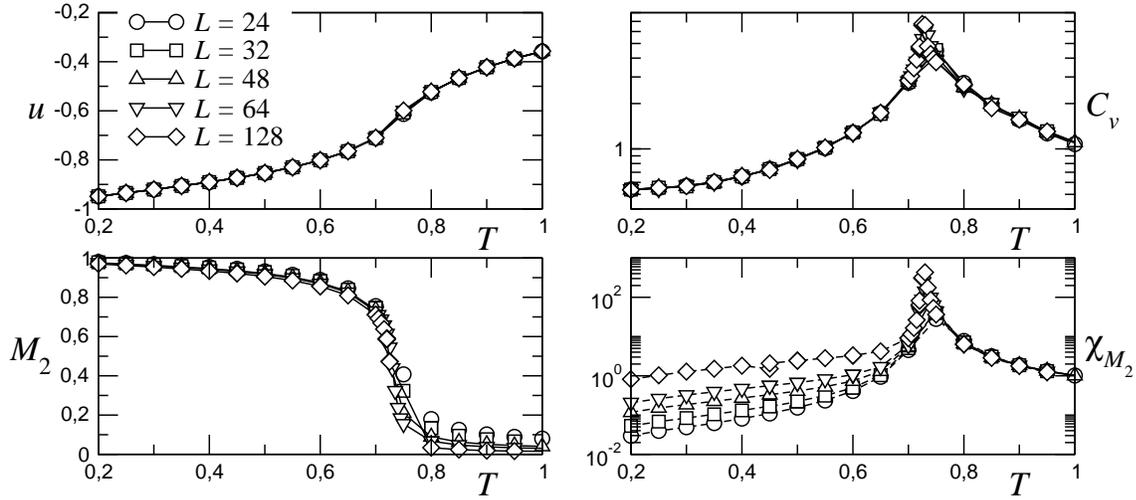}
  \end{minipage}
  \caption{Energy, specific heat, order parameter $M_2$ and corresponding 
	susceptibility $\chi_{M_2}$ vs $T$ for the $P_4\ O(2)$ model. 
	The full lines are only guides for the eyes. The values
	of $k_B$ and $J$ have been fixed to unity.}
	\label{Fig1}
\end{figure}

Simulations are performed using a standard Wolff algorithm adapted to the
expression of the nearest neighbour interaction~\cite{PalRoy03,Wolff89}. 
The spins are located on the 
vertices of a simple square lattice of size $L^2$ with periodic boundary 
conditions in the two directions. Usually $10^6$ equilibrium steps 
were used (measured as the number of flipped Wolff clusters)
and $10^6$ Monte Carlo steps for the evaluation
of thermal averages 
(the autocorrelation time at $k_BT/ J=0.2$, $L=16$
is of order of $30$~MCS, hence the numbers of MC 
steps corresponds roughly to
$3.10^4$ independent measurements for the smallest size).
A first qualitative description of the behaviour of the system is provided by 
the temperature dependence of thermodynamic 
quantities~\cite{Farinasetal05}.
The specific heat has a maximum which does not seem to  
increase substantially with the 
system size. This might be the sign of an essential 
singularity
around a temperature $k_BT_c/ J\simeq 0.70$.
From the order parameter variation, 
a smooth transition is suspected, since there is
no evolution toward a sharp jump. 
The susceptibility displays a non conventional 
behaviour at low temperature, increasing with the system size, 
which indicates a possible topological
transition with a critical low temperature phase where the susceptibility
diverges at any temperature.

\subsection{Characterization of the low-temperature phase}
We assume  the existence of a critical 
phase at low temperatures as suggested by the temperature dependence
results. The properties of the phase transition may be studied using 
\begin{description}
\item{{\em i)}}
Standard
Finite-Size Scaling technique (FSS):
in the critical low temperature phase of a model which displays a topological
transition, the physical
quantities behave like at criticality for a second-order phase transition, 
with power law behaviours of the system size. The difference is that in the
critical phase, the critical exponents depend on the temperature and for any
temperature below the transition one
has e.g. 
\be M_2(T)\sim L^{-\frac 12\eta_{M_2}(T)}\ee 
\be\chi_{M_2}(T)\sim L^{2-\eta_{M_2}(T)}.\ee 
Here $\eta_{M_2}(T)$ denotes the correlation
function critical exponent, 
\be\<Q_2(\cos(\theta_{\r_1}-\theta_{\r_2}))\>\sim|\r_1-\r_2|^{-\eta_{M_2}(T)}.\ee
\item{{\em ii)}}
Rescaling of the density profiles (or ``Finite Shape Scaling'', FShS): 
conformally covariant density profiles or correlation 
functions are exepected at {\em any temperature below the 
transition $T_{c}$}. They transform according to
$G(w_1,w_2) =
| w'(z_{1})|^{-x_{\sigma}} | w'(z_{2})|^{-x_{\sigma}}
G(z_1,z_2)$ 
through conformal mapping $w(z)$ where 
$w$ labels the lattice sites in the transformed geometry (the 
one where the computations are really performed), while $z$ refers to 
the  infinite plane where  the two-point correlations
 take the standard power-law 
expression $G(z_1,z_2)
\sim |z_1 -z_2 |^{-\eta_{\sigma}}$), 
and $x_{\sigma} = \frac{1}{2}\eta_{\sigma}$
is the scaling dimension associated to the scaling field under consideration.
Rather than two-point correlation functions, it 
is even more convenient to work with density 
profiles $m(w)$ in a finite system with symmetry breaking fields along some 
surfaces in order to induce a non-vanishing local order parameter in 
the bulk~\cite{BercheFarinasParedes02,Berche02,BercheShchur04}. 
The density $m(w)$ will be
$M_2({\bf r})=\<Q_2(\bsigma_\r\cdot{\bf h}_{\partial\Lambda})\>$.
In the case of a square lattice $\Lambda$ of size 
$L \times L$, with fixed boundary conditions along the four 
edges $\partial \Lambda$, one expects (details may be found e.g.
in~\cite{Berche02})
\begin{eqnarray}
	m(w)& \sim&
 	[\kappa(w)]^{-\frac{1}{2} \eta_{\sigma}}
	\nonumber\\
	\kappa(w)  &=&  
	\Im{\rm m} \left[ {\rm sn} \frac{2{\rm K}w}{L} \right]  \times \left| 
	\left( 1 - {\rm sn}^{2} 
	\frac{2{\rm K}w}{L} \right) 
	\left(1 - k^{2} {\rm sn}^{2} 
	\frac{2{\rm K}w}{L} \right) \right|^{-\frac{1}{2}}.
	\label{eq:squaremapping}
\end{eqnarray}
Another conformal mapping which has been applied to many 
two-dimensional critical systems is the logarithmic transformation
$w(z)=\frac {L}{2\pi}\ln z=\frac {L}{2\pi}\ln \rho+i\frac{L\varphi}{2\pi}$
which
maps the infinite plane onto an infinitely long cylinder of perimeter $L$.
The 
correlation functions along the axis of the cylinder (let say in terms of
the variable $u=\frac {L}{2\pi}\ln \rho$)
decay exponentially 
at criticality, $G(u_1,u_2)\sim\exp[-(u_2-u_1)/\xi]$,
the correlation length 
amplitude on the strip being universal, 
$\xi=\frac{L}{\pi\eta}$~\cite{Cardy84}. 
Simulations at different temperatures in a system of size 
$10\times 10000$ were performed and the correlation function exponent
thus follows from
the linear behaviour
\be\ln\<Q_2[\cos(\theta_{u_2}-\theta_{u_1}]\>={\rm const}-\frac{\pi
	\eta_{M_2}}{L}(u_2-u_1).\ee
\end{description}

\begin{figure}[ht]
  \centering
  \begin{minipage}{\textwidth}
  \epsfig{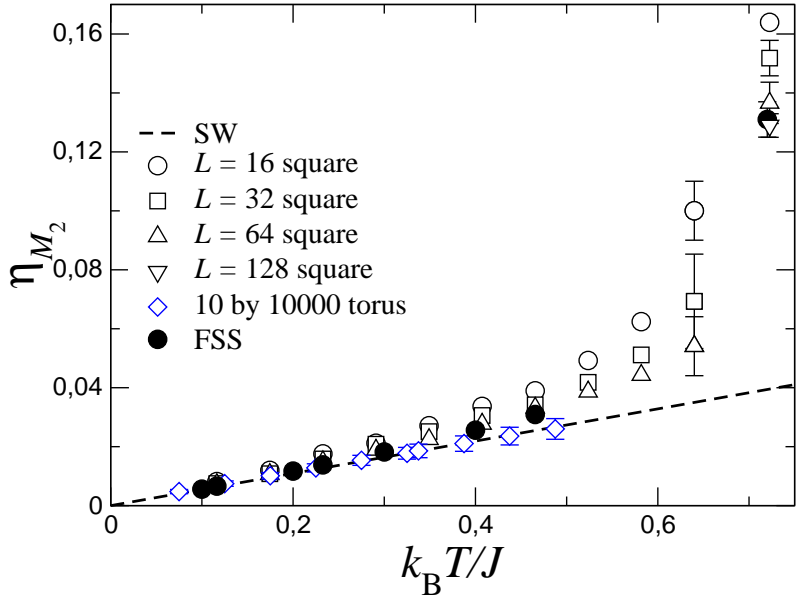}
  \end{minipage}
  \caption{Temperature variation of the correlation function exponent  
	$\eta_{M_2}(T)$ deduced from 
	conformal rescaling (open symbols) and FSS (filled symbols).
	Open triangles and diamonds, which correspond the largest systems
	seem quite reliable.  
	The dashed line shows the
	result of the spin-wave approximation
	$\eta_{M_2}(T)=\frac{11}{64}
	\frac{k_BT}{\pi J}$.
	}
	\label{Fig6and7}
\end{figure}
The resulting $\eta-$exponent deduced from these different 
techniques
is shown in Fig.~\ref{Fig6and7}. The variation of $\eta$ with the temperature
is very similar to the one observed in the case of the $XY$ model and confirms
the QLRO nature of the LT phase of the model.
The low temperature linear variation of this exponent is easily understood
within the spin wave approximation.
For $O(2)$ model with nearest neighbour interactions described by
arbitrary polynomial in $\bsigma_\r\cdot\bsigma_{\r'}$, one is led to 
an effective harmonic Hamiltonian $\frac 12J\sum_{(\r,\r')}l
(\theta_\r-\theta_{\r'})^2$.
It yields power-law correlations,
\be
	\<\cos m(\theta_{\r_1}-\theta_{\r_2})\>
	={\rm e}^{
 	-{\textstyle \frac {m^2}{2}}\langle(\theta_{\r_1}-\theta_{\r_2})^2\rangle} 
	\sim|\r_1-\r_2|^{-\eta_{ml}}
	\label{eq:SWcorrel}
\ee
with a decay exponent given by 
\be
	\eta_{ml}=\frac {m^2}{l}\eta_{XY}=
	\frac{m^2}{2\pi Kl}.
	\label{eq:eta}
\ee 
The comparison is made visible in the figure (with $m=2$ and $l=128/11$), 
and as 
expected, the lower the temperature the better the SW approximation.

\subsection{Critical behaviour at the deconfining transition}
\begin{figure}[ht]
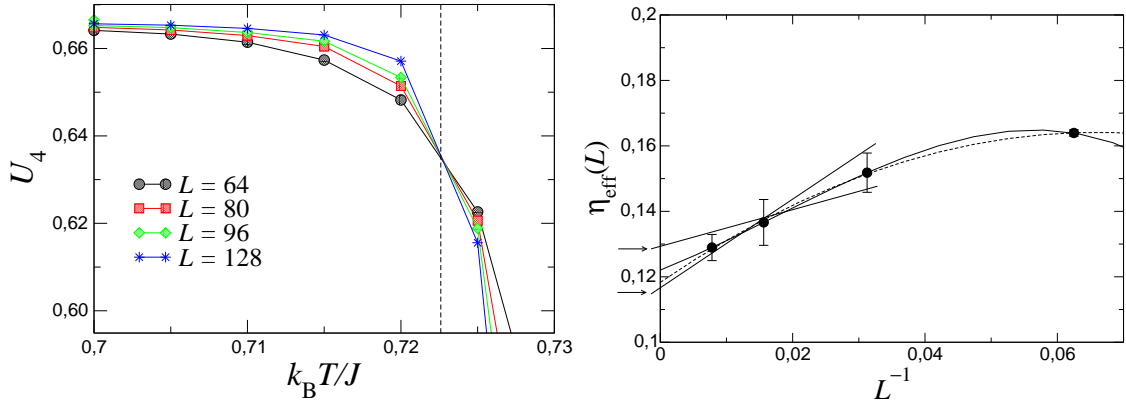

  \centering
  \begin{minipage}{\textwidth}
  \epsfig{file=Fig_U4.eps,width=0.5\textwidth}
  \epsfig{file=Fig_Eta-T_BKT.eps,width=0.48\textwidth}
  \end{minipage}
  \caption{Left: Crossing of the Binder parameter at the deconfining transition
	at a temperature
	$k_BT_{\rm BKT}/ J=0.7226$. Right: Size dependence of the
	correlation function exponent at $T_{\rm BKT}$ ($L=16$, 32, 64, 128)
	and extrapolation
	to the thermodynamic limit.}
	\label{Fig8}
\end{figure}

Not only the low temperature behaviour of $\eta$ is interesting, 
the precise value of the
$\eta$ exponent at the BKT transition where some deconfining 
mechanism should lead to the proliferation of unbinded topological defects is
also of interest, since it really describes the universality class of the
transition. An accurate value of the
transition temperature is first needed.
We performed a study of the crossing point of $U_4$ Binder cumulant for very
large statistics ($30\times 10^6$ MCS) and large system sizes (squares
of $L=64$, 80, 96 and 128 with periodic boundary conditions). The results 
shown in Fig.~\ref{Fig8} indicate a transition temperature of
$k_BT_{\rm BKT}/ J=0.7226$.

Then this temperature is used to perform Finite-Shape Scaling using
the algebraic decay of density profiles 
inside a square with fixed bounday conditions. 
These simulations are time-consuming, since the autocorrelation
time increases in the low temperature phase when $T$ evolves towards
the deconfining transition and a rather large number of Monte Carlo steps
is needed to get a satisfying number of independent measurements.
For sizes $L=16$, 32, 64, we used $10^6$ MCS for thermalization and
$30\cdot 10^6$ for measurements, while ``only'' $20\cdot 10^6$ 
for the largest size 128.
The exponential
decay of two-point correlation functions along the torus cannot be applied
at the BKT transition, 
since the system size being quite larger than in a square
geometry, the number of MC
iterations required is by far too large.
In Fig.~\ref{Fig8} we plot the ``effective'' exponent $\eta_{\rm eff}(L)$
measured at $T_{\rm BKT}$ for different system sizes as a function of
the inverse size. An estimate of the thermodynamic limit value ($L\to\infty$)
can be made using a polynomial fit (the results of  quadratic
and cubic fits are respectively $0.118$ and
$0.122$), but it is safer to keep the three largest sizes available,
$L=32$, 64, and 128, for which a linear dependence of $\eta_{\rm eff}(L)$
with $L^{-1}$ is observed. 
Taking into account the error bars, crossing the
extreme straight lines leads to 
the following value for the correlation function exponent at the
deconfining transition
\be\eta_{M_2}(T_{\rm BKT})= 0.122\pm 0.007.\ee
This value is essentially half the Kosterlitz value for the $XY$ model.

\section{Summary and open questions}\label{sec:ccl}
The results obtained are essentially the following:
\begin{description}
\item{-} The $P_4\ O(2)$ model displays a BKT-like transition with QLRO 
in the LT phase
where SWA nicely fits the nematisation temperature-dependent 
exponents $\eta(T)$
when $T\to 0$.
\item{-} The value of the exponent at the deconfining transition is supposed
to reach a universal value. The numerical estimate is close to $0.125$.
\end{description}

We believe that our conclusions 
are safe in which concerns the existence of a QLRO phase at low 
temperature. 
The results may be understood through a naive comparison 
with clock 
model in 2d. Increasing the order of the interaction polynomial indeed 
increases the number
of deep wells which stabilise the relative orientation of neighbouring spins, 
and one is thus led to a system 
which is quite similar to
a planar clock model with a finite number of states, unless the fact 
that here we keep a 
continuous spin symmetry which prevents from any ``magnetic'' 
long-range order at 
finite temperature.
The clock model is in the Potts universality class when 
$q=3$, but at 
$q\ge 4$, it displays a QLRO phase before conventional ordering at 
lower temperatures~\cite{JoseEtAl77}.
Combining this analogy  with the requirements of Mermin-Wagner-Hohenberg
theorem for continuous spin symmetry gives a natural explanation
for our results.
For any type of nearest-neighbour interaction (in $P_1$, $P_2$ or
$P_4$) the behaviour seems to be always described by a BKT-like transition.
The similar observation that a two-component nematic model renormalises in two
dimensions towards the $XY$ model was already reported in 
Ref.~\cite{NelsonPelcovits77}.
The transition is likely driven by a mechanism of condensation of defects,
like in the $XY$ model, but due to the local $Z_2$ symmetry not only
usual vortices carrying a charge $1$ are stable, but also disclination
points carrying charges $1/2$ should be stable. The role of these defects
might be studied in a similar way than in the recent work of
Dutta and Roy~\cite{DuttaRoy04}, by the comparison of the the transition
in the pure model and in a modified version where a chemical potential
is artificially introduced in order to control the presence of defects.

Now, the observed value of the exponent $\eta$ at the transition
temperature seems a bit strange. A naive application of the 
mechanism discussed in section~2 
leads to a value two times larger. 
Apart from minor modifications, the same procedure applies. 
The Hamiltonian~(\ref{eq:defHO2})
is of the form (\ref{defHKinetic}) with $K$ replaced by $Kl$ which 
does not affect the universal properties, but changes the critical 
temperature by the same factor. The other modification comes from the
local $Z_2$ symmetry in the nematic model. This changes the charges $n$
from integers to half-integers, which modifies the factor of $2\pi$ in
equation~(\ref{eqHV}) to $\pi/2$. The divergence of the
perturbative term in (\ref{eqPerturbation}) then occurs at
$\frac 12\pi K_cl=4$, where the spin wave exponent (\ref{eq:eta})
becomes
$\eta_c=m^2/16$ which takes the value $\frac 14$ when $m=2$.
We thus have a huge  discrepancy of $100~\%$ between the numerical 
determination and this prediction! A probable source of this
discrepancy may be in the mechanism involved. There are no negative charge
in the nematic model, so the pair interaction has a different
structure and a more refined analysis is therefore desirable.

\section*{Acknowledgement} 
We thank A.C.S. van Enter for useful correspondence.
Part of the material presented here (Figs. 2, 3 and 4) was 
obtained in collaboration with Ana Fari\~nas
and originally presented elsewhere. 

\section*{Note added in proofs} 
We would like to thank S. Korshunov 
who draw our attention on a possible scenario explaining the discrepancy
between the numerical value of $\eta$ and the prediction which follows from
Kosterlitz-like arguments. Twenty years ago, he studied a similar 
planar $XY$ model with mixed interactions~\cite{Koshunov86}. In the 
parameters space of the problem, there is a zone where, starting from the high 
temperature phase, a BKT transition is first observed and then at lower 
temperature occurs a phase transition governed by the presence of solitons.
This latter transition belongs to the two-dimensional Ising model (IM) 
universality class. According to this scenario, ours results would thus 
correspond to this IM transition, but the quantity that we study does
not correspond to the order parameter of this transition (hence the value
of $\eta$ is not the usual $1/4$ of the IM universality class). QLRO would
persist at higher temperatures, up to a limiting value where $\eta_{M_2}$
would indeed take its Kosterlitz-like value $m^2/16$.


\end{document}